\documentclass[prb,aps,twocolumn]{revtex4}

\usepackage{bm}
\usepackage{amsmath}
\usepackage{amssymb}
\usepackage{latexsym}
\usepackage{amsfonts}
\usepackage{epsfig}
\usepackage{psfrag}  
\usepackage{color}

\newcommand{\ket}[1]{\left|#1\right>}
\newcommand{\bra}[1]{\left<#1\right|}

\newcommand{\expval}[1]{\left< #1 \right>}
\newcommand{\dexpval}[1]{\left<\left< #1 \right>\right>}
\newcommand{\nn}{\nonumber\\}

\newcommand{\f}[1]{\mbox{\boldmath$#1$}}

\newcommand{\bea}{\begin{eqnarray}}
\newcommand{\ea}{\end{eqnarray}}
\newcommand{\eea}{\end{eqnarray}}

\newcommand{\abs}[1]{{\left| #1 \right|}}
\newcommand{\trace}[1]{{\rm Tr}\left\{ #1 \right\}}

\newcommand{\HS}{H_{\rm S}}
\newcommand{\HI}{H_{\rm SB}}

\definecolor{grey}{rgb}{0.5, 0.5, 0.5}
\definecolor{dgreen}{rgb}{0.0, 0.5, 0.0}
\definecolor{violet}{rgb}{0.5, 0.0, 0.5}
\definecolor{orange}{rgb}{1.0, 0.5, 0.0}

\begin{document}

\title{A low-dimensional detector model for full counting statistics:\\
Trajectories, Back-Action, and Fidelity}
\author{Gernot Schaller}\email{gernot.schaller@tu-berlin.de}
\author{Gerold Kie{\ss}lich}
\author{Tobias Brandes}
\affiliation{Institut f\"ur Theoretische Physik, Hardenbergstra{\ss}e 36,
Technische Universit\"at Berlin, D-10623 Berlin, Germany}
\begin{abstract}
We study the combined counting statistics of two capacitively coupled
transport channels.
In particular, we examine the conditions necessary for utilizing one channel as
detector sensitive to the occupation of the other.
A good detector fidelity may be achieved in a bistable regime when the tunneling rates
through the two channels are vastly different -- even when the
physical back-action of the detector on the probed channel is large.
Our methods allow to estimate the error of charge counting detectors
from time-resolved current measurements -- which have been
obtained in recent experiments -- alone.
\end{abstract}
\pacs{
03.65.Ta,  
05.60.Gg, 
72.10.Bg, 
73.23.Hk 
}
\maketitle


Quantum transport is a generic example of
nonequilibrium quantum dynamics~\cite{NAZ09}.
A low-dimensional quantum system (e.g. quantum dots~\cite{KOU97},
molecules~\cite{KOC05},
nanotubes~\cite{REI04} etc.) is typically coupled via particle
exchange to multiple reservoirs held at different equilibria.
The detection of stochastic particle transfer into a reservoir yields
the full counting statistics (FCS)~\cite{LEV93}.
The FCS provides a tool to access system properties via an indirect measurement.
The technique of $n$-resolved master equations~\cite{GUR96c} conveniently allows to extract the
FCS from microscopic models (see e.g.~\cite{BRA05}).
Recently, the value of this approach has been demonstrated experimentally by capacitively
coupling a single quantum dot (QD) to a quantum point contact (QPC)~\cite{GUS06}.
Here, we demonstrate that in principle, also two capacitively coupled QDs should 
show similar behavior as the QPC-QD configuration in certain parameter limits.
However, the simplicity of our model enables us to estimate the physical detector back-action 
and the detection error from the FCS.

\begin{figure}[b]
\includegraphics[width=0.47\textwidth,clip=true]{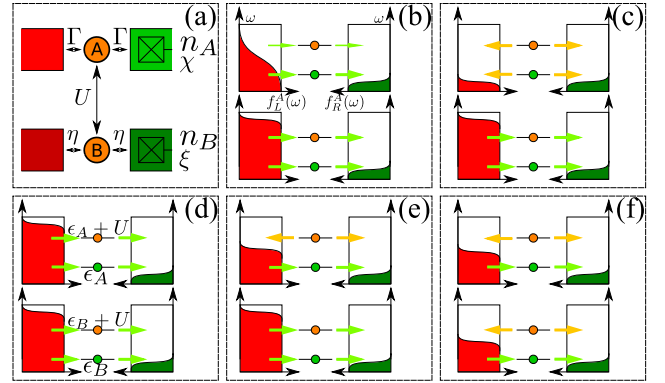}
\caption{\label{Flimits}(Color Online)
{\bf(a)} Spatial sketch of the physical setup. Tunneling is described by rates $\Gamma$ and $\eta$, respectively.
Detectors are placed at the right junctions -- with counting fields $\chi$ and $\xi$.
Coulomb interaction $U$ leads to an effective shift of the dot levels (upper orange dots in panels b-f), 
which influences the current through each channel.
{\bf(b-f)}
Bandscheme sketches with the associated Fermi functions $f_{L/R}^{A/B}(\omega)$ (labels in panel b) for different bias and 
temperature configurations (see text).
}
\end{figure}

We consider a system composed of two nearby two-terminal single resonant level systems~\cite{SAN10}
$A$ and $B$ (compare Fig.~\ref{Flimits}a)
\mbox{$\HS=\epsilon_A d_A^\dagger d_A + \epsilon_B d_B^\dagger d_B + U d_A^\dagger d_A d_B^\dagger d_B$}, 
where $\epsilon_{A/B}$ denote the level energies of the single levels and $U$ models Coulomb interaction.
The system is coupled to four fermionic reservoirs via
\mbox{$\HI=\sum_{ka} t_{ka,A} d_A c_{ka,A}^\dagger + \sum_{ka} t_{ka,B} d_B c_{ka,B}^\dagger + {\rm h.c.}$},
where $t_{ka,A/B}$ denote tunneling rates to the adjacent leads $a \in \{L,R\}$.
We will use subsystem $A$ as a detector for the state of subsystem $B$.
Transport through each level is enabled by applying nonvanishing bias voltages, and the
two transport channels influence each other by the Coulomb interaction $U$.

Since we are interested in the charge FCS not only at one junction but at the
interplay of the two channels, we introduce two virtual detectors in the tunneling terms associated with
right leads (see also Ref.~\cite{DOI07}).
Following the Born-Markov-Secular~\cite{SCH08} approximation scheme, we arrive at an
($n_A,n_B$)-resolved master equation of the form 
\bea\label{Enmresolved}
\dot \rho^{(n_A n_B)} &=& {\cal L}_{00} \rho^{(n_A n_B)}
+ {\cal L}_{+0} \rho^{(n_A-1,n_B)} + {\cal L}_{-0} \rho^{(n_A+1,n_B)}\nn
&&+ {\cal L}_{0+} \rho^{(n_A,n_B-1)} + {\cal L}_{0-} \rho^{(n_A,n_B+1)}\,,
\eea
where $\rho^{(n_A n_B)} \equiv \bra{n_A, n_B} \rho \ket{n_A, n_B}$, 
which couples only the diagonals of the density matrix to each other.
The total system (QDs and virtual detectors) density matrix can at all times be written as
$\rho(t) = \sum_{n_A, n_B} \rho^{(n_A n_B)}(t) \otimes \ket{n_A}\bra{n_A} \otimes \ket{n_B}\bra{n_B}$, 
such that the probability to measure $n_A$ tunneled particles in the detector
channel and $n_B$ tunneled particles in the system channel after time $t$ is given by
$P_{n_A n_B}(t) = \trace{\rho^{(n_A n_B)}(t)}$ (measurement postulate).
Performing a two-dimensional Fourier-transform of Eq.~(\ref{Enmresolved}) via
$\rho(\chi,\xi,t)\equiv\sum_{n_A, n_B} \rho^{(n_A n_B)}(t) e^{i n_A \chi + i n_B \xi}$, we obtain a Fourier-transformed
Liouvillian 
\mbox{$\dot\rho(\chi,\xi,t)={\cal L}(\chi,\xi) \rho(\chi,\xi,t)$}
with two counting fields
${\cal L}(\chi,\xi) \equiv {\cal L}_{00} + e^{+i \chi} {\cal L}_{+0} + e^{-i\chi}{\cal L}_{-0}
+ e^{+i\xi}{\cal L}_{0+} + e^{-i \xi}{\cal L}_{0-}$.
This equation can be solved formally, and for a given initial condition 
$\rho(0) = \rho_0^{(00)} \otimes \ket{0}\bra{0}\otimes \ket{0}\bra{0}$, 
the density matrix $\rho^{(n_A n_B)}(\Delta t)$ can be obtained from the inverse Fourier transform.
With the ''superjump'' superoperator
\bea\label{Esuperjump}
{\cal J}^{(n_A,n_B)}(t) \equiv 
\frac{1}{4\pi^2} \int_{-\pi}^{+\pi}e^{{\cal L}(\chi,\xi) t - i n_A \chi-i n_B \xi} d\chi d\xi
\eea
we obtain after time $t$ for the density matrix 
$\rho^{(n_A,n_B)}(t)={\cal J}^{(n_A,n_B)}(t) \rho_0^{(00)}$.
Measuring the number of tunneled particles through both channels at this point will collapse 
the density matrix again.
The explicit form of the Liouvillian considered in this work is given by a 
special bistable case of a size-scalable variant~\cite{SCH10} -- with an additional counting field $\xi$.

We consider only the case of unidirectional transport as shown in Figs.~\ref{Flimits}b-f:
All right-associated Fermi functions vanish at the energy scales of the system 
\mbox{$f_R^j(\epsilon_j)=f_R^j(\epsilon_j+U)=0$} (with $j \in \{A,B\}$), and the Liouvillian 
is given by ${\cal L}(\chi,\xi) = \Gamma {\cal L}_A(\chi) + \eta {\cal L}_B(\xi)$ with the 
flat tunneling rates
\mbox{$\Gamma\equiv 2\pi \sum_k \abs{t_{ka,A}}^2 \delta(\omega-\omega_{k,a})$} and
\mbox{$\eta\equiv 2\pi \sum_k \abs{t_{ka,B}}^2 \delta(\omega-\omega_{k,a})$}.
In the basis $\rho_{00,00},\rho_{10,10},\rho_{01,01},\rho_{11,11}$, the superoperators
read
\bea\label{Eliouvillian_all}
{\cal L}_A &\equiv& \left(
\begin{array}{cc}
A(\epsilon_A ) & 0\\
0 & A(\bar\epsilon_A )
\end{array}
\right),\;
{\cal L}_B \equiv \left(
\begin{array}{cc}
-B_1 & +B_2(\xi )\\
+B_1 & -B_2(0) 
\end{array}
\right),\nn
A(x)&\equiv&\left(
\begin{array}{cc}
-f_L^A(x) & e^{i\chi} + \bar{f}_L^A(x)\\
+f_L^A(x) & -1 - \bar{f}_L^A(x)
\end{array}
\right),\nn
B_1&\equiv&\rm{Diag}\big[
f_L^B(\epsilon_B),f_L^B(\bar\epsilon_B\big],\nn
B_2(\xi )&\equiv&\rm{Diag}\big[
e^{i\xi}+\bar{f}_L^B(\epsilon_B),e^{i\xi}+ \bar{f}_L^B(\bar\epsilon_B)\big],
\eea
where $\bar{f}_{L/R}^j(x)\equiv 1-f_{L/R}^j(x)$ and $\bar\epsilon_j\equiv\epsilon_j+U$.
Bistability occurs when e.g. $\eta\to 0$ and $f_L^A(\epsilon_A) \neq f_L^A(\bar\epsilon_A)$, 
since the block structure of the Liouvillian supports two different currents (and stationary states) 
in this case~\cite{SCH10}.



\begin{figure}
\includegraphics[width=0.47\textwidth,clip=true]{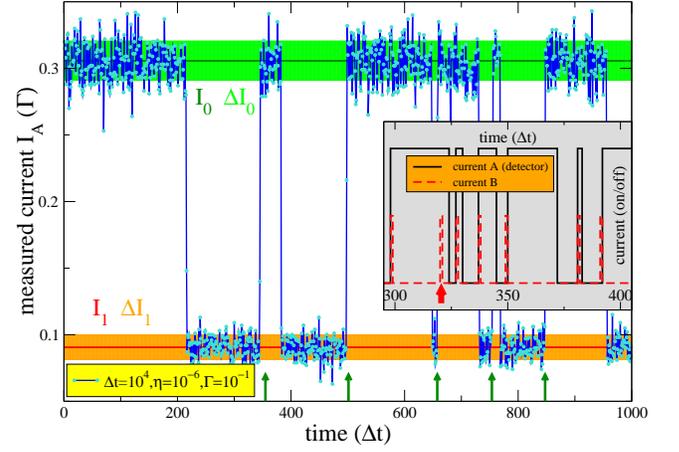}
\caption{\label{Fcvec}(Color Online)
Trajectory for configuration in Fig.~\ref{Flimits}b for 1000 current 
measurements at time intervals of $\Delta t$.
Background lines display the currents and widths (shaded regions) for the single
resonant level evaluated at the effective QD level one obtains for empty (high current)
or filled (low current) channel $B$.
Temperature and chemical potentials chosen such that $f_L(\epsilon_A)=0.612$ and $f_L(\bar\epsilon_A)=0.181$.
{\bf Inset} (parameters $\Gamma \Delta t=10$, $\eta\Delta t=0.1$, $\Delta t=100.0$ arb. u.) shows the 
correlation of both (coarse-grained) currents for the configuration in Fig.~\ref{Flimits}e.
When the detector current (solid black) rises, one most often also
observes a spike (single charge) in the current through channel $B$ (dashed red, rescaled).
However, due to stochastic behavior of the detector, jump events may be missed completely (bold red arrow).
}
\end{figure}

{\bf\em Current Trajectories.}
When channel $B$ is at infinite bias and at sufficiently 
large temperatures $k_{\rm B} T \gtrsim U$, the left-associated Fermi functions
$f_L^A(\epsilon_A)$ and $f_L^A(\bar\epsilon_A)$ will assume intermediate values between zero and one
(see Fig \ref{Flimits}b),
which yields two different currents through A depending on whether $B$ is occupied or not.
Ignoring the number of transferred charges through channel $B$, the probability of obtaining $n_A$ charges
after time $\Delta t$ equates to
$P_{n_A}(\Delta t)={\rm Tr}\{\sum_{n_B} {\cal J}^{(n_A,n_B)} (\Delta t) \rho_0\}
\equiv {\rm Tr}\{ {\cal J}^{(n_A)}_A(\Delta t)\rho_0\}$.
That is, to obtain a trajectory of current measurements performed equidistantly at intervals $\Delta t$ numerically, 
a random number according to the distribution $P_{n_A}(\Delta t)$ must be drawn.
The outcome of this then corresponds to a measurement of $n_A$ particles
(with current $I_A \equiv n_A/\Delta t$).
For simplicity, we exploit the translational invariance of Eq.~(\ref{Enmresolved}) and shift $n_A$ to zero
after each measurement.
However, for the subsequent evolution, one now has to
use the normalized density matrix $\rho^{(n_A)}$ as initial condition, which leads to
a temporal correlation of the measured currents, see Fig.~\ref{Fcvec}.
In principle, also for unidirectional transport the number of different 
outcomes ${\mathfrak M} \in \{0,1,2,\ldots\}$ is infinite, but fortunately, there exists a natural
cutoff as for $n_A \ge \Gamma \Delta t$ the operators ${\cal J}^{(n_A)}_A(\Delta t)$ become exponentially small.
The trajectory in Fig.~\ref{Fcvec} is very similar to results obtained for QPCs~\cite{GUS06}.
The blips from low to high currents (arrows) should therefore 
indicate single charges tunneling from channel $B$ to its right junction.
However, as tunneling through channel $A$ is also a stochastic process, jump events may in principle be
missed (compare also the inset).
Therefore, this raises the question of detector fidelity and back-action.
In the following, we will investigate this for configurations (c)-(f) in Fig.~\ref{Flimits}.

%

{\bf\em Unperturbed Level (Fig.~\ref{Flimits}c).} 
In this limit, the current through $A$ is blocked completely, such that the detector is turned off.
With the replacements $f_L^A(\epsilon_A)=f_L^A(\bar\epsilon_A)=0$ and
$f_L^B(\epsilon_B)=f_L^B(\bar\epsilon_B)=1$ in Eq.~(\ref{Eliouvillian_all}),
we obtain for the Laplace transform
$\tilde M(\chi,\xi,z) \equiv \trace{\left[z\f{1}-{\cal L}(\chi,\xi)\right]^{-1} \bar \rho}$
of the moment-generating function (MGF)
\bea\label{Emgfinfbias}
\tilde M(\chi,\xi,z)=\tilde M_B(\xi,z) = \frac{\left(3+e^{i \xi}\right) \eta+2z}{2\left[z^2+2\eta z-\left(e^{i \xi}-1\right)\eta^2\right]}\,,
\eea
from which one may analytically~\cite{SCH10} deduce the probability distribution
of a single resonant level in the symmetric infinite bias limit.


{\bf\em Infinite Bias (Fig.~\ref{Flimits}d).}
When both channels are held at infinite bias -- replacements $f_L^A(\epsilon_A)=f_L^A(\bar\epsilon_A)=1$ and
$f_L^B(\epsilon_B)=f_L^B(\bar\epsilon_B)=1$ in Eq.~(\ref{Eliouvillian_all}) -- 
the detector cannot distinguish the different states of the probed system. 
From the exact matrix exponential of the Liouvillian we obtain 
$P_{n_A n_B}(t) = P_{n_A}(t) P_{n_B}(t)$, where each probability distribution corresponds to that
of a single resonant level~\cite{SCH10} (with $\eta\to\Gamma$ for channel $A$).
In Laplace space, this becomes visible by considering the reduction to Eq.~(\ref{Emgfinfbias})
via $\tilde M(0,\xi,z) = \tilde M_B(\xi,z)$ and similarly 
$\tilde M(\chi,0,z) = \tilde M_A(\chi,z)$.
Clearly, we obtain for the first and second (long-term limit) cumulants of the detector 
(current and noise) 
$\dexpval{n_A}=\frac{\Gamma t}{2}$ and $\dexpval{n_A^2}\to \frac{1}{8}+\frac{\Gamma t}{4}$ and for the probed system
$\dexpval{n_B}=\frac{\eta t}{2}$ and $\dexpval{n_B^2}\to \frac{1}{8}+\frac{\eta t}{4}$.
From the factorization of the probability distributions it follows also that the
cross-correlations vanish at all times $\dexpval{n_A n_B}\equiv\expval{n_A n_B}-\expval{n_A}\expval{n_B}=0$.


{\bf\em Charge Detector Limit (Fig.~\ref{Flimits}e).}
When $f_L^A(\epsilon_A)=1$, $f_L^A(\bar\epsilon_A)=0$, and
$f_L^B(\epsilon_B)=f_L^B(\bar\epsilon_B)=1$ in Eq.~(\ref{Eliouvillian_all}) -- the detector
is bistable~\cite{SCH10}, which can be used to distinguish the two different states of channel $B$:
When channel $B$ is occupied, current through $A$ is blocked completely, whereas it is maximal (at its infinite
bias value) otherwise.
We find that the Laplace transform of the MGF fulfills
$\tilde M(0,\xi,z) = \tilde M_B(\xi,z)$ -- compare Eq.~(\ref{Emgfinfbias}) --
which demonstrates that here the physical action of the detector on the probed system vanishes.
In contrast, the physical influence of the system on the detector is not negligible, which is demonstrated
by $\tilde M(\chi,0,z) \neq \tilde M_A(\chi,z)$.
Therefore, the role of the detector will be purely informational (measurement postulate) without 
physical back-action on the system.

We obtain for the first cumulants 
$\dexpval{n_A}=\frac{\Gamma t}{4}$, $\dexpval{n_B}=\frac{\eta t}{2}$, whereas the second cumulants asymptotically approach
the long-term limits
\bea
\dexpval{n_A^2} &\to& \frac{1}{32} \left(1 - \frac{\Gamma^2}{\eta^2}\right) 
+ \frac{3 \eta^2 + 3 \eta \Gamma + \Gamma^2}{16 \eta (\eta + \Gamma)} \Gamma t\,,\nn
\dexpval{n_B^2} &\to& \frac{1}{8}+\frac{\eta}{4} t\,,\qquad
\dexpval{n_A n_B} \to  -\frac{\Gamma}{16 (\eta + \Gamma)}\,,
\eea
i.e., there exists a slight negative cross-correlation between the two channels, which surprisingly
saturates.

In order to operate the channel $A$ as a detector for the FCS of channel $B$, 
we need to scan channel $A$ at much larger rates than $B$ ($\Gamma \gg \eta$).
In particular, to obtain a meaningful detector current 
[associated with a transiently bimodal~\cite{SCH10} distribution $P_{n_A}(\Delta t)$], 
we require $\Gamma \Delta t \gg 1$ and to see the slow switching in time-resolved current measurements 
it is also necessary that $\eta \Delta t \ll 1$.
Then, measuring a vanishing detector current $I_A$ within a time interval $\Delta t$ indicates with high
probability that system $B$ is occupied, and measuring a non-vanishing detector current indicates an empty channel 
$B$.

To investigate within which limits on $\Gamma$, $\eta$, and $\Delta t$ this simplistic view
is valid, we introduce the coarse-grained superoperators 
${\cal J}^{LL}\equiv{\cal J}^{(00)}(\Delta t)$,
${\cal J}^{LH}\equiv\sum_{n_B\ge 1} {\cal J}^{(0 n_B)}(\Delta t)$,
${\cal J}^{HL}\equiv\sum_{n_A\ge 1} {\cal J}^{(n_A 0)}(\Delta t)$, and 
${\cal J}^{HH}\equiv\sum_{n_A,n_B \ge 1} {\cal J}^{(n_A n_B)}(\Delta t)$,
corresponding to a vanishing ($L$) or non-vanishing ($H$) current through channels $A$ and $B$.
That is, the $L/H$-discrimination threshold is set here independent of $\Delta t$.
From Eq.~(\ref{Esuperjump}), we may analytically obtain their Laplace transforms
by performing the summation in frequency space (not shown).
We may then create coarse-grained current trajectories with both detectors $A$ and $B$ 
(leading to four different measurement outcomes 
${\mathfrak M} \in \{LL, LH, HL, HH\}$), see inset of Fig.~\ref{Fcvec}.
Obviously, the correlation between blips of current $I_A$ and spikes of current $I_B$ is not perfect:
The detector result may fail, as e.~g. tunneling charges
through channel $B$ may be missed.

\begin{figure}[t]
\includegraphics[width=0.47\textwidth,clip=true]{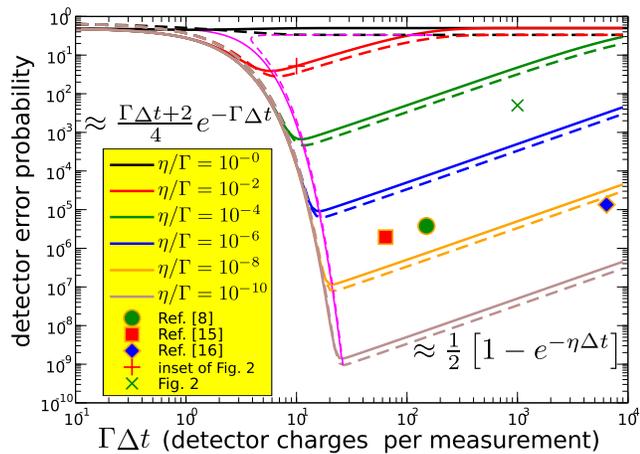}
\caption{\label{Ferrorprob}(Color Online)
Detector error probability in the quantum detector limit [Eq.~(\ref{Eperrideal}), solid lines] and in the 
dynamical channel blockade limit [Eq.~(\ref{Eperrdcb}), dashed lines] for varying ratios of average tunneled 
detector and system charges $\eta/\Gamma$.
For a too small or too large number of tunneled detector charges, the detector result is
practically useless, whereas for an intermediate number, 
a pronounced minimum (note the logarithmic plot) of the error probability
is found.
Thin curves in the background mark the trajectory of the minimum.
Asymptotic formula refer to solid lines (c) -- dashed curve asymptotics 
would have slightly different prefactors.
Filled symbols yield rough error estimates for experiments.
}
\end{figure}

Without the FCS of channel $B$, the error probability of the detector can be calculated from the averaged 
joint probabilities of measuring a large current in channel $A$ and 
a particle in channel $B$ and measuring a vanishing current through $A$ in combination with 
no particle in $B$:
$P_{\rm err}^A(\Delta t)\equiv{\rm Tr}\{d_B^\dagger d_B {\cal J}^{H}_A \bar \rho\}
+ {\rm Tr}\{(\f{1}-d_B^\dagger d_B) {\cal J}^{L}_A \bar \rho\}$, 
where $\bar\rho$ is defined via ${\cal L}(0,0) \bar\rho=0$,
${\cal J}^{H}_A\equiv{\cal J}^{HL}+{\cal J}^{HH}$, and 
${\cal J}^{L}_A\equiv{\cal J}^{LL}+{\cal J}^{LH}$.
Its Laplace transform equates to
\bea\label{Eperrideal}
\tilde P_{\rm err}^A(z) = 
\frac{1}{2 z} - \frac{\Gamma^2 \left[2 (2 \Gamma + z)^2 + \eta (7 \Gamma + 4 z)\right]}
{8 (\Gamma+\eta) {\cal P}_{\rm num}(z)}
\eea
with 
${\cal P}_{\rm num}(z) \equiv 2 \eta \Gamma^2 (\eta + \Gamma)\linebreak[1] 
+ 2 \Gamma (\eta + \Gamma) (4 \eta + \Gamma) z\linebreak[1]
+ (2 \eta + \Gamma) (2 \eta + 5 \Gamma) z^2\linebreak[1]
+ 4 (\eta + \Gamma) z^3\linebreak[1]
+ z^4$.
From the Laplace transform we can directly conclude that for too
short or too long measurement times the detector result will be useless:
$\lim_{\Delta t\to 0} P_{\rm err}^A(\Delta t) = \frac{1}{2}$ and 
$\lim_{\Delta t\to \infty} P_{\rm err}^A(\Delta t) = \frac{1}{2}$.
An intuitive explanation This can be understood intuitively: For too short measurement times, the two peaks of the corresponding
bimodal distribution will not have well separated, and measuring a vanishing particle number (current) 
$n_A(\Delta t)$ is still possible even when channel $B$ is empty.
In contrast, for too large measurement times, we may obtain an average current 
over several cycles of loaded and empty channel $B$.
For intermediate measurement times $\Delta t$ however, we observe a pronounced minimum of $P_{\rm err}^A(\Delta t)$, 
see Fig.~\ref{Ferrorprob}.


{\bf\em Dynamical Channel Blockade (Fig.~\ref{Flimits}f).}
Here, both probed system ($B$) and detector ($A$) may be seen as bistable systems~\cite{SCH10}, 
depending on whether $\eta$ or $\Gamma$ dominate, respectively.
When occupied, both transport channels block the current through the other one completely \cite{COT04}.
With the replacements $f_L^A(\epsilon_A)=1$, $f_L^A(\bar\epsilon_A)=0$, $f_L^B(\epsilon_B)=1$, 
and $f_L^B(\bar\epsilon_B)=0$ in Eq.~(\ref{Eliouvillian_all}), it becomes obvious 
that the doubly charged state $\ket{11}$ will not be occupied.
When we choose the junctions to $A$ and $B$ and the corresponding tunneling rates identically 
($\eta \to \Gamma$ and $\xi \to \chi$), we recover the formal Liouvillian structure in Ref.~\cite{BEL05}
for a two-terminal two-level system.
The Laplace transform of the MGF $\tilde M(\chi,\xi,z)$ may also be obtained analytically, and we find that 
the interactions and back-actions between system and detector are not negligible, exemplified by
$\tilde M(0,\xi,z) \neq \tilde M_B(\xi,z)$ and $\tilde M(\chi,0,z) \neq \tilde M_A(\chi,z)$.
From the MGF we obtain for the first cumulants $\dexpval{n_A}=\frac{\Gamma t}{3}$, $\dexpval{n_B}=\frac{\eta t}{3}$,
and the second cumulants approach the asymptotic limits
\bea
\dexpval{n_A^2} &\to& \frac{8 \eta^2 - 2 \eta \Gamma - 4 \Gamma^2}{81 \eta^2}
+\frac{5 \eta + 2 \Gamma}{27 \eta} \Gamma t\,,\nn
\dexpval{n_B^2} &\to& 
\frac{8\Gamma^2-2\eta \Gamma - 4 \eta^2}{81 \Gamma^2}
+ \frac{2 \eta + 5 \Gamma}{27 \Gamma} \eta t\,,\nn
\dexpval{n_A n_B} &\to& 
+2 \frac{\eta^2 - \eta \Gamma + \Gamma^2}{81 \eta \Gamma}
-\frac{\eta+\Gamma}{27} t\,,
\eea
such that the cross-correlations do now also appear in the currents.
We may also here introduce the superoperators corresponding to either zero or more than zero particles
leaving either transport channel, and the current trajectories (not shown) are similar to the inset of 
Fig.~\ref{Fcvec}.
The Laplace transform $P_{\rm err}^A(\Delta t)$ is obtained similarly
\bea\label{Eperrdcb}
\tilde P_{\rm err}^A(z) &=& 
\frac{z^2 (3 \Gamma + 2 z) + \eta (\Gamma^2 + 6 \Gamma z + 4 z^2)}
{3 z (\Gamma + z) \left[z (\Gamma + z) + \eta (\Gamma + 2 z)\right]}\,,
\eea
and we can directly conclude that for too
short or too long measurement times the detector result will be useless since
$\lim_{\Delta t\to 0} P_{\rm err}^A(\Delta t) = \frac{2}{3}$ and 
$\lim_{\Delta t\to \infty} P_{\rm err}^A(\Delta t) = \frac{1}{3}$.
For finite measurement times the error probability will show a pronounced minimum, 
see Fig.~\ref{Ferrorprob}.
Therefore, measurements on $A$ may give reliable information on the state of $B$ 
(which differs strongly from its uncoupled dynamics)


{\bf\em Detector Errors and Experiments.}
Eqns.~(\ref{Eperrideal}) and~(\ref{Eperrdcb}) generate a map of detector fidelity (Fig.~\ref{Ferrorprob}).
The detection error only depends on the two
dimensionless variables $\Gamma \Delta t$ (number of detector charges during the measurement time)
and $\eta/\Gamma$ (ratio of system to detector currents) and can be made extremely small.
The striking similarity of the detector error probability for ideal and non-ideal detectors
(also the similarity of the current trajectory in Fig.~\ref{Fcvec} with experiments 
for QPCs \cite{GUS06}) indicates that the simple dependence on detector and system 
charge throughput is generic for $I_{\rm det}^{\rm min} \ll I_{\rm det}^{\rm max}$.
An error estimate taking into account a realistic $\Delta t$-scaling of the $I_{\rm low}/I_{\rm high}$ discrimination
threshold, asymmetric and energy-dependent tunneling rates, multiple levels etc. 
would be much more accurate, but for a crude estimate we may link
experimental parameters with those of our model:
The width of the experimental current divided by its mean value will in the high-bias limit
yield the average number of tunneled detector charges during the measurement time via
$\Delta I_{\rm det}^{\rm max}/I_{\rm det}^{\rm max} \approx 1/\sqrt{\Gamma \Delta t}$, and the
ratio of mean system current (number of blips divided by time) and mean detector current yields
$I_{\rm sys}/I_{\rm det}^{\rm max} \approx \eta/\Gamma$, 
and we obtain error probabilities of approximately $10^{-5}$ per measurement for Refs.~\cite{GUS06,FRI07,SUK07}.



{\bf\em Conclusions.} 
Even with strong physical back-action 
the detector may yield reliable results for the instantaneous occupation of channel $B$ (with an
altered dynamics).
To decide whether the used detector strongly back-acts on the probed system, 
it is therefore more useful to observe the scaling of the channel
cross correlations: The smaller the scaling coefficient of the cross-cumulant in time, the
smaller is the detector back-action.
In order to estimate the fidelity of a single current measurement from Fig.~\ref{Ferrorprob}, 
we propose to use the characteristics of time-resolved experimental detector current trajectories.

Financial support by the DFG (project \mbox{BR 1528/5-1}) and discussions with 
R. Sanchez are acknowledged.



\end{document}